\documentclass{aa}
\usepackage{epsf}
\begin{document}


\def\asec{^{\prime\prime}}  \def\mic{\,\mu\rm m} 
\def\csi{\,{\rm cm}^{-2}}           
\def\cci{\,{\rm cm}^{-3}}          
\def\fu{{\rm W}\,{\rm cm}^{-2}\,\mic^{-1}}    
\def\fum{{\rm W}\,{\rm m}^{-2}\,\mic^{-1}}
\def\lu{{\rm W}\,{\rm cm}^{-2}}         
\def\lum{{\rm W}\,{\rm m}^{-2}}
\def\Teff{T_{\rm eff}}              \def\Tbb{T_{\rm BB}}
\def\kms{\rm km\,s^{-1}} 

\def\Hi{H{\sc\,i}}       
\def\Hii{H{\sc\,ii}}     
\def\HH{$H_2$}            
\def\hei{He\sc\,i\rm}       
\def\arii{[Ar{\sc\,ii}]}    \def\ariii{[Ar{\sc\,iii}]}   
\def\neii{[Ne{\sc\,ii}]}    \def\neiii{[Ne{\sc\,iii}]}   
\def\Niii{[Ni{\sc\,ii}]}
\def\siii{[S\sc\,iii\rm]}   \def\siv{[S{\sc\,iv}]}
\def\fFeii{[Fe{\sc\,ii}]}   \def\fFeiii{[Fe{\sc\,iii}]}  
\def\pFeii{Fe{\sc\,ii}}     
\def\Siii{[Si{\sc\,ii}]}
\def\Bra{Br-$\alpha$}   \def\Brb{Br-$\beta$}  \def\Brg{Br-$\gamma$} 
\def\Paa{Pa-$\alpha$}
\def\Lsun{L_\odot}         \def\Msun{M_\odot}
\def\Av{A_V}               \def\Ak{A_K}
\def\Asil{A_{\rm sil}}     \def\tausil{\tau_{\rm sil}}

\def\lesssim{\;\raise.4ex\hbox{$<$}\kern-0.8em \lower .6ex\hbox{$\sim$}\;}
\def\moresim{\;\raise.4ex\hbox{$>$}\kern-0.8em \lower .6ex\hbox{$\sim$}\;}
\def\tbc{\bf TBC\rm}

\def\plothalf#1{\centering \leavevmode\epsfxsize=.45\textwidth\epsfbox{#1}}
\def\plotone#1{\centering \leavevmode \epsfxsize=\textwidth \epsfbox{#1}}
\def\plottwo#1#2{\centering \leavevmode\epsfxsize=.48\textwidth 
    \epsfbox{#1} \hfil\epsfxsize=.48\textwidth \epsfbox{#2}}

\thesaurus{08
           (08.03.4; 08.23.2; 09.04.1; 13.09.6)
	  }

\title {Mid-infrared imaging
 and spectroscopy of the enigmatic cocoon stars in the Quintuplet
 Cluster\thanks{ISO is an ESA project with instruments funded by ESA Member
 States (especially the PI countries: France, Germany, the Netherlands and
 the United Kingdom) and with the participation of ISAS and NASA.}
       }

\author{
   A.~Moneti\inst{1} \and
   S.~Stolovy\inst{2} \and
   J.A.D.L.~Blommaert\inst{3} \and
   D.F.~Figer\inst{4} \and 
   F.~Najarro\inst{5} 
       }

\offprints{A. Moneti}
\titlerunning{The Cocoon Stars in the Quintuplet Cluster}

\institute{Institut d'Astrophysique, 98bis Blvd. Arago, F-75014 Paris,
    France \\ email: moneti@iap.fr 
\and
    CalTech, Astronomy Dept., 105-24,  Pasadena, CA 91125, USA \\
    email: srs@astro.caltech.edu
\and
    ISO Data Centre, P.O.~Box 50727, E-28080 Madrid, Spain \\
    email: blommaert@iso.vilspa.esa.es
\and
    Space Telescope Science Institute, 3700 San Martin Drive,
    Baltimore, MD 21218, USA \\ email: figer@stsci.edu
\and
Instituto Estructura de la Materia, CSIC, Serrano 121, E-29006, 
    Madrid, Spain \\ email: najarro@isis.iem.csic.es}

\date{Received .../ accepted}
\maketitle


\begin{abstract}

In an attempt to determine the nature of the enigmatic cocoon stars in the
Quintuplet Cluster, we have obtained mid-infrared imaging and
spectrophotometry of the cluster, using the CAM and SWS instruments on ISO,
using SpectroCam-10 on the Palomar 5m telescope, and NICMOS on HST.  The
spectra show smooth continua with various dust and ice absorption features.
These features are all consistent with an interstellar origin, and there is
no clear evidence for any circumstellar contribution to these features.  We
find no spectral line or feature that could elucidate the nature of these
sources.  Detailed modeling of the silicate absorption features shows that
they are best reproduced by the $\mu$ Cep profile, which is typical of the
interstellar medium, with $\tausil \simeq 2.9$.  The high spatial
resolution mid-IR images show that three of the five cocoon stars have
spatially extended and asymmetric envelopes, with diameters of $\sim
20,000\,$AUs.

A reddening law similar to that of Lutz (1999) but with silicate features
based on the $\mu$ Cep profile and normalized to our value of $\tausil$ is
used to deredden the observed spectrophotometry.  The dereddened energy
distributions are characterised by temperatures of 750--925 K, somewhat
cooler than determined from near IR data alone.  Models of optically thin
and geometrically thick dust shells, as used by Williams et al.~(1987) for
very dusty, late-type WC stars, reproduce the observed SEDs from 4 to
$17\mic$, and imply shell luminosities of $\log (L/\Lsun) \simeq 4.5$--4.9
for the brightest four components.  An analysis of the various suggestions
proposed to explain the nature of the cocoon stars reveals serious problems
with all the hypotheses, and the nature of these sources remains an enigma.

\end{abstract}



\section{Introduction}

The Quintuplet Cluster (AFGL 2004) is one of the three young clusters known
in the vicinity of the Galactic Centre.  It was first identified in the
near infrared survey of Glass, Catchpole, \&\ Whitelock (1987) as a bright
source coincident with AFGL 2004, and found to be a bright IRAS source with
a cool energy distribution (Glass 1988).  It is located about 30 pc from
the Centre in projection, and its cluster nature became apparent with the
advent of IR cameras through the works of Glass, Moneti, \& Moorwood (1990;
GMM90), Nagata et al.~(1990), Okuda et al.~(1990), Moneti, Glass, \&
Moorwood (1992), and Moneti, Glass, \&\ Moorwood (1994 [MGM94]), who
identified over a dozen stars in the cluster, including five with
extraordinarily large luminosities ($\sim 10^5 \Lsun$), cool (600---1200 K)
energy distribution, and featureless $K$-band spectra.  These
characteristics led those authors to tentatively identify them as massive,
dust-enshrouded young stars, henceforth \em cocoon \rm stars.  The
possibility that they be OH/IR stars was discarded since (i) they have no
CO bandhead at 2.3 and $4.6\mic$, (ii) they show little or no variability
(Glass et al.~1999 find peak-to-peak of $\sim 0.45 \pm 0.05\,$mag and K for
two of them, no variability for the remainder), (iii) they lack OH maser
emission, (iv) they are cooler and more luminous than typical OH/IR stars,
and (v) they are much younger than OH/IR stars, assuming they are coeval
with the cluster ($\approx 4$ Myr, Figer McLean and Morris 1999 [FMM99],
Figer et al.~1999).  More recently, FMM99 suggested that the cocoon stars
might be extremely dusty, late-type WC stars (DWCL, Williams et al.~1987,
WHT), even though they do not detect the expected emission features in the
$J$-band.  And finally, Glass et al.~(1999) suggested they could be
self-obscured, very massive O stars close to the ZAMS, which, according to
Bernasconi and Maeder (1996), go through a prolonged accretion phase of
2--2.5 Myr, comparable to the age of the Quintuplet Cluster, before
becoming visible.

In an attempt to elucidate the nature of the cocoon stars, we have obtained
spectroscopy and spectrophotometry in the mid-IR, where these sources are
brightest, in order to (i) search for spectral features intrinsic to the
stars, (ii) obtain accurate measurements of their spectral energy
distribution, (iii) to investigate their spatial extension, and (iv) to
study the shape of the silicate feature in particular to determine whether
it could be partly intrinsic to the sources.


\section{Observations, calibration, and results}

Two instruments onboard the Infrared Space Observatory (ISO, Kessler et
al.~1996) were used: ISOCAM (Cesarsky et al.~1996) was used to obtain high
spatial and low spectral resolution Circular Variable Filter (CVF)
spectrophotometry of the Quintuplet Cluster from 2.5 to $17\mic$, and SWS
(de Graauw et al.~1996) was used to obtain two high resolution spectra at
two positions in the cluster.  Table 1 lists the ISO identification number
(ION) of the datasets used.  Of these, the last was proposed and planned by
the authors, while the remainder were obtained by other investigators and
taken from the archive.  Some preliminary results based on the ION 09901781
dataset have already presented Nagata et al.~(1996).  Since that time, the
calibrations have improved considerably and, in particular, the CVF
spectral response function used here is based on stellar observations
obtained in orbit, and differs considerably from the one used in Nagata et
al.~(1996), which was obtained on the ground prior to launch.

We also present high spatial resolution ground-based 8--$14\mic$
spectroscopy and narrow-band imaging photometry of the cluster members,
obtained with SpectroCam-10 on the Palomar 5 m telescope, and 1--$2.5\mic$
photometry derived from HST/NICMOS images.  These data are particularly
useful to understand the limits of low spatial resolution ISO data.

\begin{figure}         
\plothalf{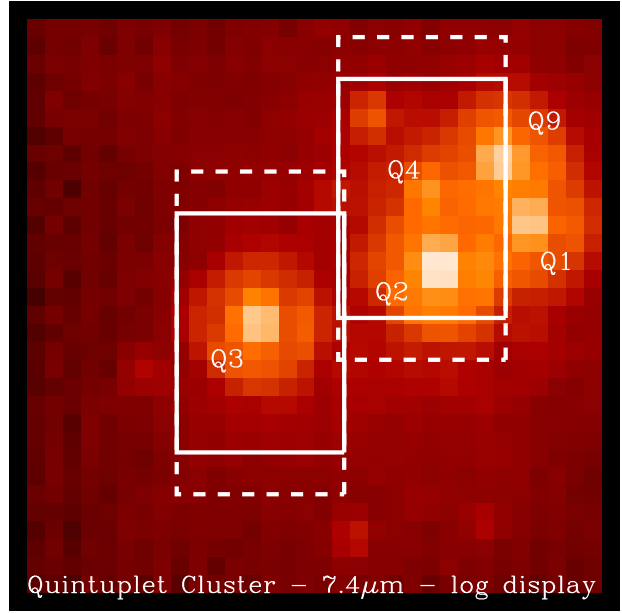}
\caption{Location of the SWS apertures on a $7.4\mic$ ISOCAM image of the
Quintuplet Cluster.  The intensity scale is logarithmic, north is up and
east is to the left.  The naming convention is that of GMM90. Solid line is
aperture 1 ($14''\times 20''$), and dashed lines is aperture 2 ($14''\times
27''$).  The source to the NE of Q4 was denoted Q5 in GMM90.}
\end{figure}

\subsection{SWS Grating Spectroscopy}

\begin{table}
\caption{ISO Datasets} 
\leavevmode \footnotesize
\begin{tabular}{lcccc}  
\hline \\ [-5pt]
ISO ION  & Instr. & mode & date \\ [2pt]
\hline \\ [-5pt]
09901781 & CAM & C04 (CVF)   & 24-Feb-96 \\
28701246 & SWS & S01-speed 3 & 29-Aug-96 \\
29702147 & SWS & S01-speed 3 & 09-Sep-96 \\
84500303 & CAM & C04 (CVF)   & 09-Mar-98 \\ [-2pt]
\hline 
\end{tabular} 
\end{table}

The SWS spectra were obtained in the AOT1-speed 3 mode, which scans the
grating continuously over its full range, and produces a number of spectral
segments, corresponding to the various aperture/grating/detector
combinations.  These segments cover the full spectral range of
2.3---$47\mic$ with a nominal resolving power ranging between 250 and 600.
In this work we will be concerned primarily with data out to $28\mic$,
which was obtained with the two smallest apertures, which are nominally
$14''\times 20''$, $14''\times 27''$, respectively, though in practice
slightly larger [A. Salama, private communication].  The aperture changes
at $12\mic$.  The location of the nominal apertures on a $7.4\mic$ images
CVF image is shown in Figure 1, where the cocoon stars are identified
following the naming convention of Moneti, Glass, \&\ Moorwood (1992).  The
SWS apertures were centered on Q4 and Q3 (respectively GCS 3-I and GCS 4 in
Kobayashi et al.~1983).

\begin{figure*}        
\centering \leavevmode
\epsfxsize=0.83\textwidth\epsfbox{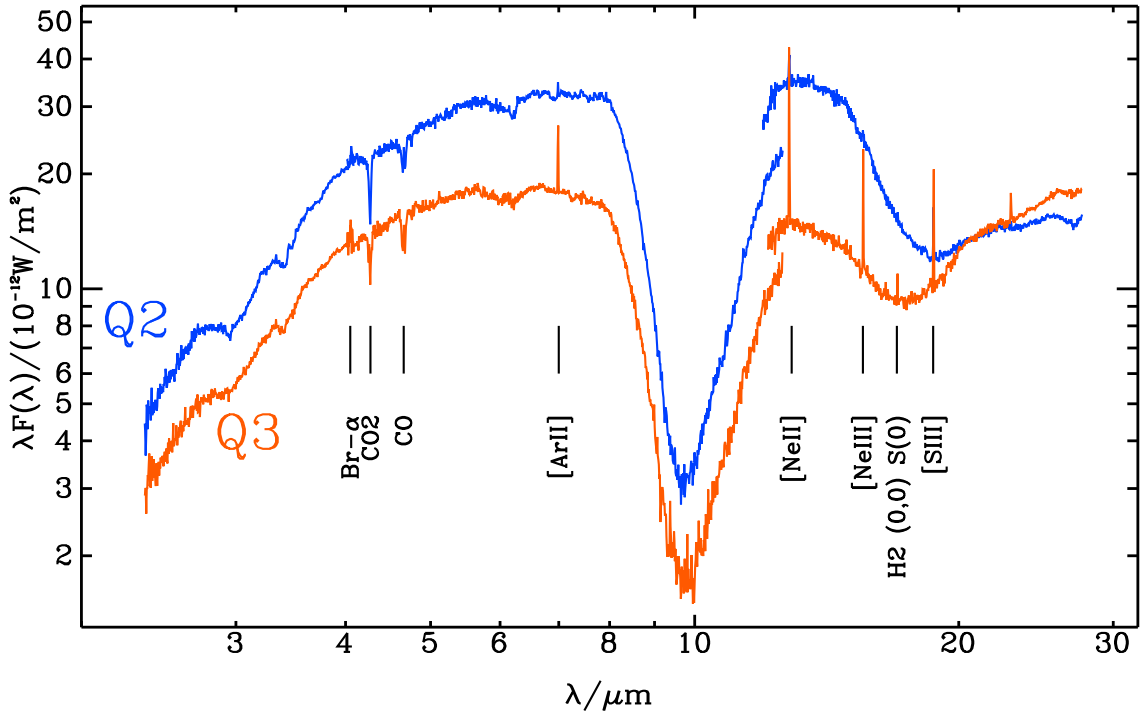}
\caption{SWS spectra of Q2 (primarily) and Q3.  The jump at $12\mic$
corresponds to the aperture change; see text for details.  }
\end{figure*}

Data processed with version 7.0.1 of the ISO Pipeline were further cleaned
with ISAP\footnote{The ISO Spectral Analysis Package (ISAP) is a joint
development by the LWS and SWS Instrument Teams and Data Centers.
Contributing institutes are CESR, IAS, IPAC, MPE, RAL and SRON.} by
removing data from noisy detectors, and, for each spectral segment, the
data from the different detectors were normalized to the average of the
complete set of data, effectively compensating for slight residuals in dark
current (at low signals) or in detector responsivity (at high signals).
Following these steps, the agreement between adjacent segments was quite
good; the results are shown in Figure 2.  The jump at $12\mic$ is due to
the change in effective aperture, and is expected when observing extended
sources.  At wavelengths beyond $28\mic$ the effective SWS apertures are
even larger and they include a significant part of the Pistol \Hii\ region
whose hot dust emission dominates the spectrum.  These parts of the spectra
will not be considered further.

The calibration used in the SWS Pipeline assumes that the target is
point-like and in the center of the aperture.  In that case, the nominal
overall uncertainties would be $\sim 5\%$ up to $4.1\mic$, $\sim 10\%$ for
4.1---$27\mic$, and $\sim30\%$ beyond $29\mic$ (see de Graauw et al.~1999
for details).  This is essentially the case for Q3, which is sufficiently
isolated from the other sources.  In the case of the GCS-3 complex, which
includes Q1, 2, 4, and 9, the aperture was centered on Q4, but the
brightest source in the beam is Q2, located $6''$ south of Q4, and
significant light is scattered into the aperture from the other components,
especially at long wavelengths, where the PSF is broadest.


\begin{figure*}            
\centering \leavevmode
\epsfxsize=\textwidth\epsfbox{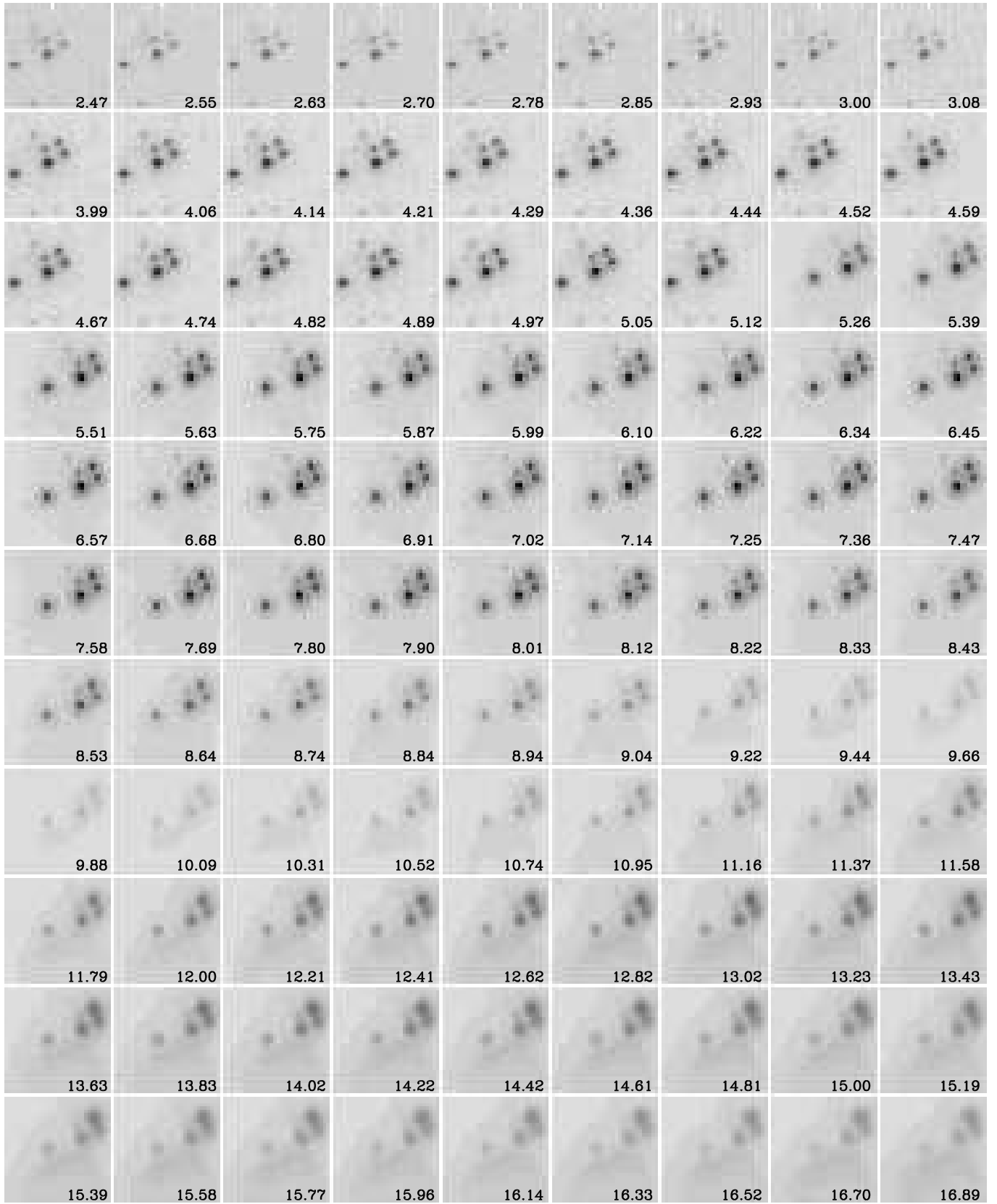}
\caption{Composite of most of the CVF images of the Quintuplet Cluster.
Wavelengths in $\mic$ are labeled.  Individual panels are normalized in
$\lambda F(\lambda)$ units. North is up and east is to the left, and each
panel is $48''$ on the side.  The jump in image position between 5.12 and
$5.26\mic$ is due to the switch from the SW to the LW section of ISOCAM.}
\end{figure*}

\subsection {CAM CVF Imaging}

The ISOCAM instrument, used to obtain CVF (Continuously Variable Filter)
scans, contains two separate cameras that cover the regions shortward and
longward of $5\mic$ (henceforth `SW' and `LW', respectively).  Both cameras
utilize $32\times 32$ pixel arrays, an InSb array with CID readout for the
SW channel and a Si:Ga array with a direct read-out for the LW channel.
The spectral resolving power ranges from 45 to 100.  For each CVF segment,
a single scan was obtained with wavelength increasing with time, and each
scan was preceded by a long (40---100) series of frames obtained at the
first CVF position for detector stabilisation.

We present here 2.5--$3.0\mic$ and 4--$5\mic$ spectra from the ISO 09901781
dataset, and 5--$17\mic$ spectra from the ISO 84500303 dataset.  All these
data were obtained with a pixel scale of $1.5''$.  The ISO 09901781
dataset also contains 5--$17\mic$ data obtained with the 3 arcsec/pixel
scale, but these data are, for the most part, heavily saturated, and are
not discussed here.  In both cases, the total integration time per spectral
position was approximately 1 min.

For completeness we also examined the ISO 49400231 and the ISO 49400268
datasets, which contain $3\times 3$ rasters of the cluster at two CVF
positions and with two broad filters.  The filter data suffer from
significant saturation, and the CVF data are qualitatively of much lower
quality than the complete CVF scans, probably as a results of transient
effects during the raster scanning process.

The ISOCAM data were reduced and calibrated with the IDL-based CAM
Interactive Analysis (CIA) package\footnote{CIA is a joint development by
the ESA Astrophysics Division and the ISOCAM consortium, led by the ISOCAM
PI, C.~Cesarsky, Direction des Sciences de la Mati\`ere, C.E.A., France}
following the standard steps of (i) dark subtraction, (ii) removal of
cosmic ray events, (iii) correction of the behaviour of the detector in
response to signal jumps (transient correction, LW only, using the
``inversion'' method), (iv) averaging all valid images obtained at a given
CVF position, (v) flatfielding the results using library flats, and (vi)
conversion the average signal to physical units using standard conversion
factors, originally derived from several standard stars.  Throughout the
analysis, the dead column (\#24) in the LW array was replaced by the
average of the adjacent columns.  No ghost image could be identified in the
CVF data, nevertheless some light could be lost to low-level, diffuse ghost
images.  From a study of the background, sampled in the NE corner of the
array, see \S 3.2, we find that the background level is 2\% of the
total signal in the array at $\lambda < 6\mic$, then rises quickly to 5\%
at $8\mic$, and remains at that level thereafter.

A composite of the resulting data is shown in Figure 3, where the
individual panels are normalized in $\lambda F(\lambda)$ units.  Note in
particular (i) the nearly complete silicate absorption, (ii) that the
brightest source at short wavelength, Q2, is no longer so at long
wavelength, where Q9 becomes brightest, and (iii) the increase in PSF size
with wavelength.

The nebula south of the cluster, which becomes prominent at $\lambda >
10\mic$, is the Pistol nebula, and the Pistol star is prominent at the
bottom of the frames at the short wavelengths.  These are the subject of a
separate paper (Moneti et al.~2000, in preparation).

The main difference between these results and those of Nagata et al.~(1996)
is that the spectral response function for the SW-CVF section ($\lambda <
4\mic$) has changed considerably (Biviano et al.~1997), so that the new
flux densities in the 2.5--$3\mic$ range are now lower by almost a factor
of 2, and the $3\mic$ ice feature is shallower than previously reported.
As will be shown below, the new results are in better agreement with
previously published results.


\subsection{Palomar Imaging and Spectroscopy}

High spatial resolution (1$''$) images and low-resolution ($R\sim 50$)
spectra were obtained in the 8--$13\mic$ atmospheric window with
SpectroCam-10 (Hayward et al.~1993) on the Palomar 5m Hale telescope.
These observations were taken on July 10--12, 1993 (spectroscopy and
$11.7\mic$ imaging) and June 14, 1995 ($8.7\mic$ imaging). SpectroCam-10
uses a $128 \times 128$ Rockwell Si:As detector.  Its camera mode has a
fully illuminated field of view of $\sim 15''$ diameter with $0.256''$
pixels.  All observations employed the standard mid-infrared techniques of
chopping the secondary mirror (typically at 5-10 Hz) and nodding the
telescope (typically $60''$ E-W, every few minutes) in order to adequately
subtract the thermal background.  Images were obtained with two standard
OCLI filters centered at 8.74 and $11.66 \mic$ with half-power bandwidths
of $\Delta\lambda=0.78$ and $1.11\mic$, respectively, and mosaics were
constructed from these images.  The flux calibrators were $\alpha$ Boo and
$\alpha$ Her, and their fluxes were obtained from Hanner et al.~(1984).  An
airmass correction was derived from observations of standard stars at
different airmasses; the cocoon stars were observed at a typical airmass of
2.3.

\begin{figure}       
\plothalf{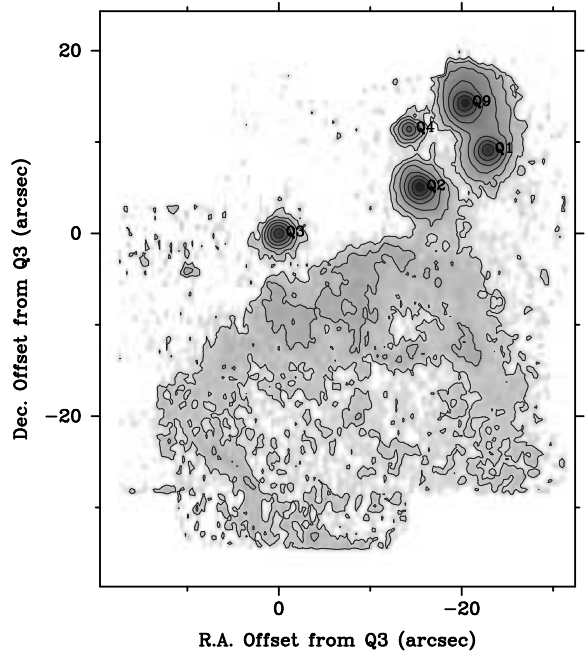}
\caption{Mosaic of the Quintuplet Cluster and the Pistol Nebula at
$11.7\mic$.  Pixels are $0.256''$ wide, the grey-scale is logarithmic to
show the large dynamic range, and the contour levels are 1.3, 3.1, 8.0, 120
49, 123, and $311\times 10^{-14}\fum\,{\rm arcsec}^{-2}$ (0.05, 0.12, 0.31,
0.76, 1.9, 4.8, and 12.1 Jy$\,{\rm arcsec}^{-2}$). }
\end{figure}

Figure 4 represents a mosaic of 47 seeing-selected, flat-fielded $11.7\mic$
images, each with a total on-source integration time of 40 seconds. During
the mosaic process the pixels were subsampled by factor of three in order
to optimise image registration; after mosaicking, the images were rebinned
back to the original pixel sizes, and the result was smoothed slightly with
a Gaussian of $\sigma = 0.8$ pixel.  A mask was used to omit pixels not
fully illuminated.  The image registration accuracy in the region
containing the cocoon stars is estimated to be better than 0.3 pixles, or
$\lesssim 0.1''$; in regions without bright stars, the telescope offsets
were used, and the registration accuracy is only $\sim 1''$.  The $8.7\mic$
image was constructed from 56 images with 5 seconds on-source integration
time each.

The spectroscopy was obtained with a $2'' \times 15''$ slit to yield a
resolution of $\lambda/\Delta \lambda \sim 50$.  The slit was oriented N-S,
the dispersion was $0.046\mic$/pixel, and four pixels were combined for each
resolution element.  Four grating settings were used to cover to
8--$13\mic$ range and each setting was exposed for a total integration time
on-source of 20 seconds for each of the cocoon stars.  Atmospheric lines
were used for wavelength calibration and $\beta$ Peg was observed for flux
calibration and to correct for atmospheric lines.


\begin{figure*}[bht]       
\centering \leavevmode 
\epsfxsize=0.95\textwidth\epsfbox{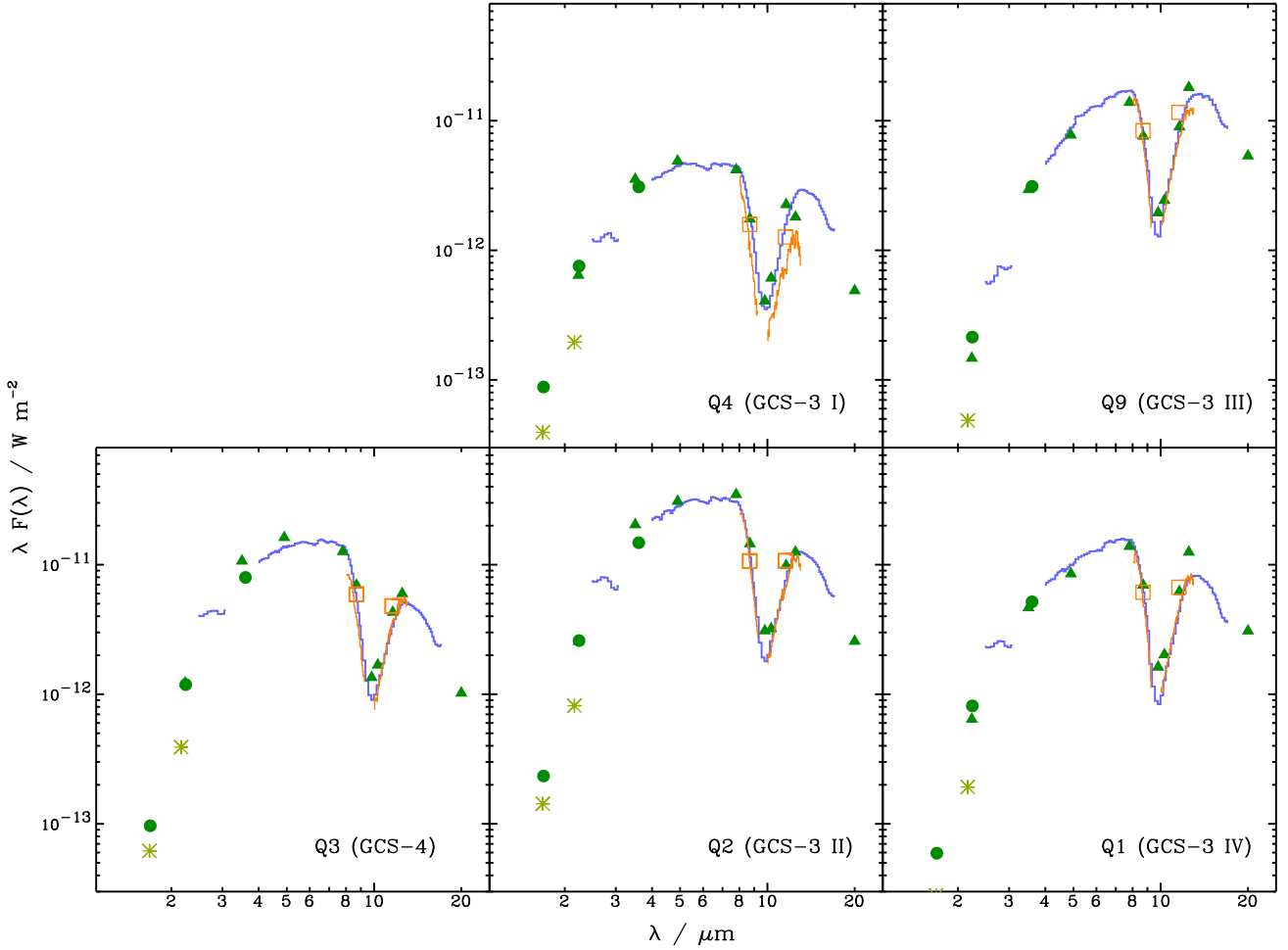}
\caption{Observed spectra and photometry of the cocoon stars.  The J-band
data and some H-band data were left off scale in order to avoid overly
compressing the ordinate, which would make the spectral features too small
to see (these data points are plotted in Figure 11 below).  The CVF spectra
are the blue histogram, Palomar spectra and photometry are orange histogram
and squares, NICMOS photometry are green asterisks.  The green circles are
near-IR imaging photometry from FMM99, and the green triangles are mid-IR
aperture photometry from Okuda et al.~(1990).  The panels show the sources
roughly in their positions on the sky. }
\end{figure*}

\subsection{HST/NICMOS Imaging}

The Quintuplet cluster was observed with NICMOS on HST in program 7364 in
Sept. 1997 with Camera 2, giving a spatial resolution of $0.075''$/pix.
Three broadband filters were used: F110W, F160W, and F205W.  Figer et
al.~(1999c) give details of the observations.  The raw data were retrieved
from the STScI archive and were re-reduced with an improved linearity
correction algorithm using the `calnica' task in STSDAS.  This had the
effect of recovering photometric information for the central pixels of the
Quintuplet stars, which were flagged as saturated in the pipeline data
reduction.  Mosaics were constructed by aligning 16 dithers in each filter
to produce the final images.  Residual dc levels were measured and
subtracted from the images.


\subsection{Photometric and Spectrophotometric Results}

Photometry of the cocoon stars was extracted from each CVF image by
integrating within a small aperture and multiplying the result by an
aperture correction factor that was determined from CVF observations of the
standard star $\delta$ Draconis.  The sky background at each wavelength was
determined from regions of the image free of stars, and the same sky level
was used for all sources.  Even with a small aperture the photometry of the
faintest sources can suffer significant contamination from the wings of the
brighter sources nearby.  This effect is most serious at long wavelength
where the PSF is largest.  This problem is most acute for Q4, which is the
weakest source and which is surrounded by three bright ones.

\begin{table*}
\caption{Photometry (in magnitudes) of the cocoon stars}
\leavevmode  \footnotesize
\begin{tabular}{llrrrrrr}
\hline \\ [-5pt]
Source  & GCS &  [F110W] & [F160W] & [F205W] & [8.7] & [11.7] \\
\hline \\ [-5pt]
Q1 & 3-IV & 16.63 & 11.48 & 8.37 & 1.14 &  0.13 \\
Q2 & 3-II & 14.46 &  9.71 & 6.80 & 0.54 &--0.38 \\
Q3 & 4    & 15.31 & 10.62 & 7.60 & 1.18 &  0.49 \\
Q4 & 3-I  & 15.78 & 11.11 & 8.35 & 2.61 &  1.94 \\
Q9 & 3-III& 18.88 & 13.54 & 9.86 & 0.80 &--0.46 \\
\hline \\ [-5pt]
$\lambda_{\rm iso}/\mic$ & & 1.14 & 1.63 & 2.10 & 8.69 & 11.55 \\
$A_\lambda/\Av$  & & 0.285 & 0.177 & 0.119 & 0.073 & 0.081 \\
FZM$_\lambda$/Jy & & 1571 &  1075 & 695 & 50.4 & 28.9 \\
\hline 
\end{tabular} \end{table*}

The SpectroCam-10 photometry was extracted with a 5--$7''$ (diameter)
aperture, depending on the extendedness of the source.  The NICMOS
photometry was derived from the energy enclosed in a 3-pixel radius
($0.45''$ diameter) aperture, and then corrected to an infinite aperture
using measured isolated stars to derive the aperture correction.  These
correction factors were 1.31, 1.50, and 1.71 for the F110W, F160W, and
F205W filters, respectively.  The results are listed in Table 2.  We assign
these values conservative uncertainties of 0.2 mag, which are due primarily
to the photometric extraction method and to errors in the estimate of the
residual dc levels.  The actual uncertainty for the brighter sources is
likely to be closer to 0.1 mag, especially at the longer wavelengths.

Conversion to flux densities are based on the system in which Vega (without
its mid-IR excess) is assigned 0-mag in all bands (Cohen et
al.~1992)\footnote{The Vega model used for the absolute calibration can be
found in
www.iso.vilspa.esa.es/users/expl\_lib/ISO/wwwcal/\-iso\-prep/\-cohen/\-vega9605.dat.}
While conversion of magnitudes to flux densities is straightforward, the
determination of the proper wavelength at which the flux density should be
plotted is more complicated, since it depends on the shape of the
(observed) energy distribution of the source.  For this purpose, we have
modeled the intrinsic shape of the 1--$4\mic$ energy distributions as warm
blackbodies, and reddened them with the reddening law discussed below to
produce a model {\em observed} energy distribution.  This was then used to
compute the isophotal wavelength, $\lambda_{\rm iso}$, as defined by the
relation \[ F(\lambda_{\rm iso}) \int T(\lambda) d\lambda = \int F(\lambda)
T(\lambda)d\lambda, \] where $F(\lambda)$ is the observed spectrum of the
source, and $T(\lambda)$ is the transmission of the instrument (including
filter, detector efficiency, optics, and possibly the earth's atmosphere).
For $\lambda_{\rm iso}$ to be unique, $F(\lambda)$ must be a monotonic
function of $\lambda$, which is indeed the case for our models.  The
isophotal wavelengths for the NICMOS and SpectroCam-10 filters are listed
in Table 2 together with the values of reddening correction
($A_\lambda/A_V$) and the adopted flux density for a zero magnitude
FZM$_\lambda$/Jy.

Figure 5 shows the observational results, which are compared with the
ground-based photometry of FMM96 (filled circles) and of Okuda et
al. (1990, filled triangles, see also corrections in Nagata et al.~1996).
Approximate $\lambda_{\rm iso}$ values were computed for the FMM99 data
using standard J, H, and K filter profiles, but not for the mid-IR data,
which was obtained with narrow-band filters.

The agreement between the CVF and the Palomar spectra is excellent, as is
the agreement between the CVF spectra and the ground-based photometry.  The
discrepancy between the CVF and the Palomar spectra of Q4 at $\lambda
\moresim 10\mic$ is due to the contamination of the CVF spectrophotometry
of Q4 by its bright neighbours (see above).  We consider the CVF spectrum
unreliable there, and will instead rely on the Palomar results.

At NIR wavelengths, the NICMOS fluxes fall below the ground-based results
since the former can better isolate the target sources from other nearby
sources (see Figure 2 of Figer et al.~1999c)

\subsection{ Detailed comparison of SWS and CVF spectra}

A comparison of the CVF and the SWS spectra is presented in Figure 6.  The
SWS spectra were transformed to surface brightness using the measured
aperture sizes, which are $1.12\times$ and $1.29\times$, and $1.06\times$
larger than the nominal apertures for Bands 1, 2, and 3, respectively
(A. Salama, private communication).  These factors, derived from beam
profiles, are 0th order approximations of a parameter that is a function of
wavelength.  The change of correction factor introduces a $\sim 10\%$ jump
at the Band 1/Band 2 interface ($4\mic$), where the nominal aperture
remains unchanged.  The CAM spectra were extracted using a synthetic
rectangular apertures approximating the {\sl measured} apertures of SWS,
and centered as shown in Figure 1.

\begin{figure*}    
\centering \leavevmode 
\epsfxsize=0.83\textwidth \epsfbox{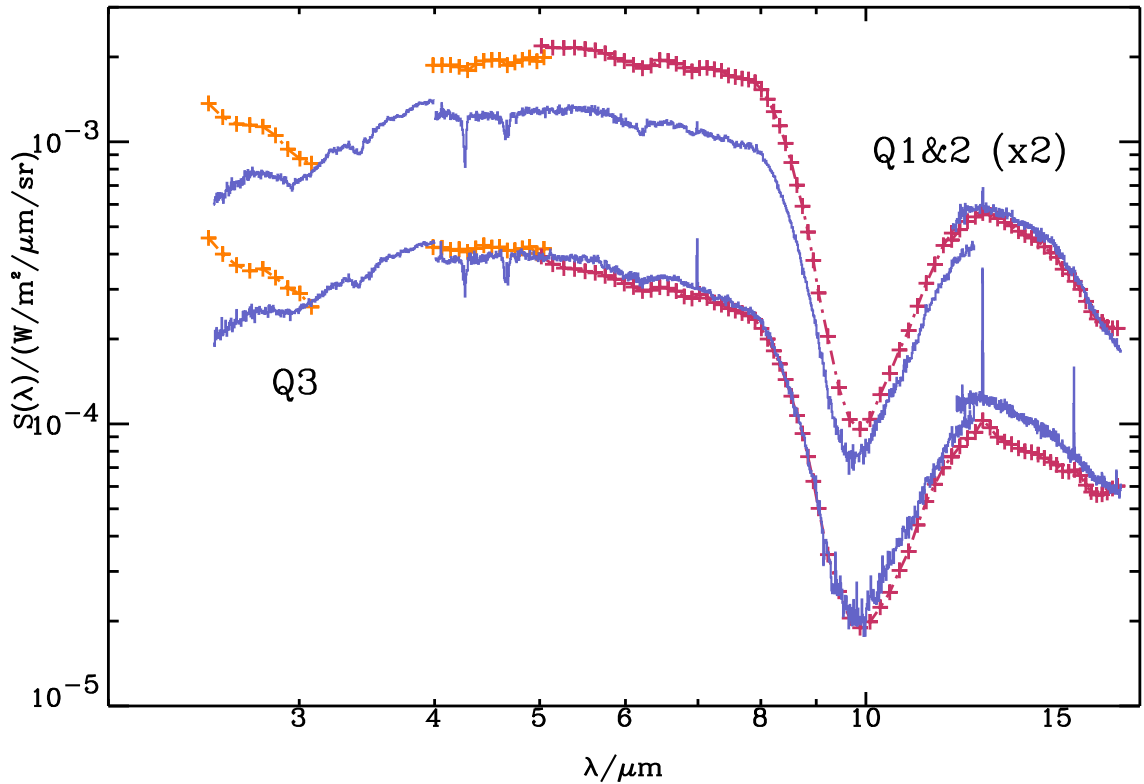}
\caption{Comparison of SWS spectra (bluw histogram) with spectra extracted
from CVF images (red and orange crosses, indicating different CVF
sections).}
\end{figure*}

Over the 4--$12\mic$ range, SWS Band 2, the agreement between the SWS and
the CVF spectra is excellent for Q3, which is fairly isolated.  The SWS
spectrum of Q2 falls short of the CVF result; this difference is due to
the fact that Q2 is near the edge of the SWS beam, where the responsivity
is much lower than in the beam centre.

For Band 3, the SWS shows an increase in surface brightness for both
spectra.  No such increase is expected for Q3 on the basis of the CVF
spectra, while a minor one is expected for Q2.  The difference is likely to
come from the detailed shape of the SWS beam profile, which is not
accounted for in this simulation.

In Band 1 the CVF spectra show a rapid decrease with wavelength, which is
unexpected and contrary to the SWS behaviour.  We note that this region
corresponds the the beginning of the CVF spectrum, where detector transient
effects are most serious (no transient correction was applied to the SW
data since the behaviour of these detectors was not sufficiently well
characterised), and to the edge of the SW-CVF, where the determination of
the response function is most uncertain.  We consider this portion of the
CVF spectra unreliable.

We conclude that, with the exception of the first part of the CVF spectrum,
the agreement between the two datasets is excellent, and this suggests that
the overall uncertainties could be less that $\sim 10\%$ over most of the
spectral range.

\section{Results}

\subsection{Emission lines}

One of the objectives of the SWS spectra of the cocoon stars was to search
for emission lines that could clarify the nature of these stars.  A careful
search of the spectra has revealed only ionized hydrogen and forbidden
lines attributable to the Pistol \Hii\ region.  These are stronger in the
spectrum of Q3, which is closer to the \Hii\ region so that more line (and
continuum) emission is included in the SWS aperture.  In contrast, only the
strongest lines are detected in the Q2 spectrum, and these are detected
mostly at the longest wavelengths, where the SWS aperture is largest.
Furthermore, these lines are not detected in the CVF and Palomar spectra,
albeit at lower resolution, confirming that they are not intrinsic to the
cocoon stars.  In contrast, these lines are clearly detected in the CVF and
Palomar spectra of the Pistol \Hii\ region.  Beyond $28\mic$ the spectra
(not shown) were obtained with even larger apertures, and the contamination
by the Pistol Nebula, both in terms of forbidden line emission and of
thermal emission from the warm dust (Moneti et al.~2000), begins to
dominate over the cocoon stars.  Thus, no emission line attributable to the
cocoon stars themselves was detected.


\subsection{Dust and Gas Features}

The spectra contain many dust absorption bands; these have been the subject
of a number of recent studies (Schutte et al.~1998, Gerakines et al.~1999,
Chiar et al.~2000), while Moneti and Cernicharo (2000) have studied the
gaseous components (all these results come from the same SWS spectra used
here).  These results show that the lines of sight toward Q2 and Q3 contain
(i) cold molecular cloud material, traced by absorption by cold ($\simeq
10\,$K) CO and H$_2$O gas and by various ice species (H$_2$O, CO$_2$, CO,
and CH$_4$), and more tenuous ISM material, traced by aliphatic and
aromatic hydrocarbon features.  Special attention has been paid to the
narrow absorption feature at $6.2\mic$, attributed to the C--C stretch in
aromatic hydrocarbons.  Schutte et al.~(1998) suggested that it could be
the UIR feature seen in absorption, and justify the lack of absorption at
3.28 and $11.3\mic$ on excitation and temperature grounds, while they find
that the observed spectrum is consistent with the presence of some
absorption at $7.7\mic$.  The best detection of this absorption feature is
toward WR118, a DWCL star whose absorption spectrum is believed to due to
diffuse ISM material only, and marginal detections are reported towards
three other DWCL stars (Schutte et al.~1998).  Chiar et al.~(2000), noting
the $6.2\mic$ absorption in the cocoon stars, which might also be DWCL
stars, suggested that this absorption feature could be associated with the
environment of the DWCL stars rather than to the diffuse ISM.  However,
Chiar et al.~(2000) also note that WC stars are not likely to be
significant contributors to the carrier(s) of the C--H stretch ($3.28\mic$
band), so that if the 3.28 and $6.2\mic$ bands (and the other UIR bands)
are indeed connected, it is unlikely that they arise primarily in the
surroundings of late-type WC stars.

\begin{figure}    
\plothalf{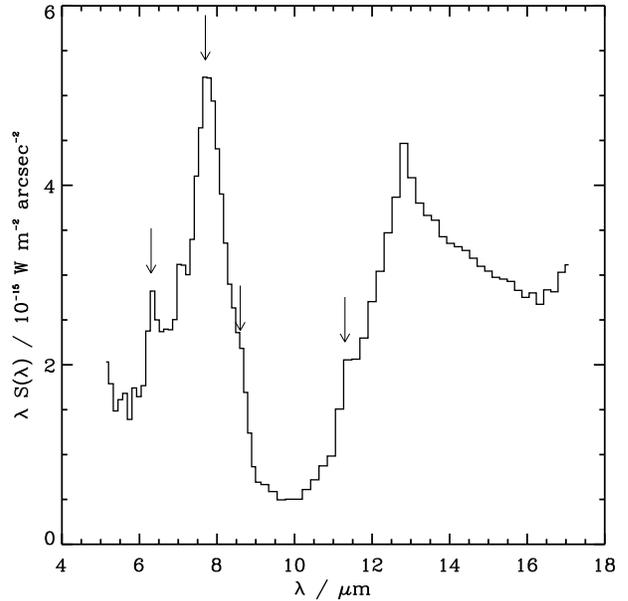}
\caption{The UIR features as seen in the spectrum of the background, in
this case a $5\times 5$ pixel region in the north-east corner of the array.
The UIR features at 6.2, 7.7, 8.6, and $11.3\mic$ are identified by the
arrows.}
\end{figure}

In contrast to the cocoon stars, the CVF data clearly show the UIR bands
in emission in the spectrum of the background, as shown in Figure 7.  The
intensity of this emission is only $\sim 2$ and 3\% of the continuum of Q2
and Q3, respectively, which is weak compared to the measured absorption.
The background emission also shows the forbidden lines of \arii\ and \neii,
and an important level of continuum.  A part of this emission could be due
to light scattered within the instrument.

We have attempted to locate the source of the $6.2\mic$ emission from the
CVF data by computing the difference of the line planes of the datacube
minus the adjacent continuum planes.  The result was flat, indicating that
the emission is uniform over the whole region (unlike the \arii\ and \neii\
emission, which occur in the Pistol \Hii\ region).  We deduce that the UIR
emission is a characteristic of the line of sight rather than of the cocoon
sources.  We cannot determine where along the line of sight the UIR
emission originate on the basis of our data.

\subsubsection{The Silicate Features}

The silicate absorption bands at 9.7 and $18\mic$ are the strongest
features in the cocoon star spectra.  We have used several published
emissivity profiles to model the observed absorption.  Most published
profiles concern the $9.7\mic$ feature which is observable from the ground:
Roche and Aitken (1984, RA84), Hanner, Brooke and Tokunaga (1995, HBT95),
Bowey, Adamson, and Whittet (1998, BAW98), while two determinations include
both features: P\'egouri\'e and Papoular (1995) and Simpson (1991).  The
studies of RA84 and BAW98 are based on absorption of ISM dust, HBT95 study
the absorption by molecular cloud dust, and the remainder are based on the
dust emission around late type supergiant stars, which are generally
believed to reproduce the ISM profile.  These profiles, normalized to unity
at their peak value, are shown in Figure 8.  A detailed comparison of these
emissivities is beyond the scope of this work, but it is worth noting that
there are important differences, even among those that should be very
similar (e.g.~the RA84 $\mu$ Cep profile and the Simpson 1991 group 1
profile, which is based on several stars with narrow $9.7\mic$ features,
including $\mu$ Cep itself).  Some of the differences should probably be
ascribed to the different methods of determining and modeling the
continuum, which will produce important differences in the wings of the
profiles.

\begin{figure*}[bht]    
\centering \leavevmode 
\epsfxsize=0.83\textwidth\epsfbox{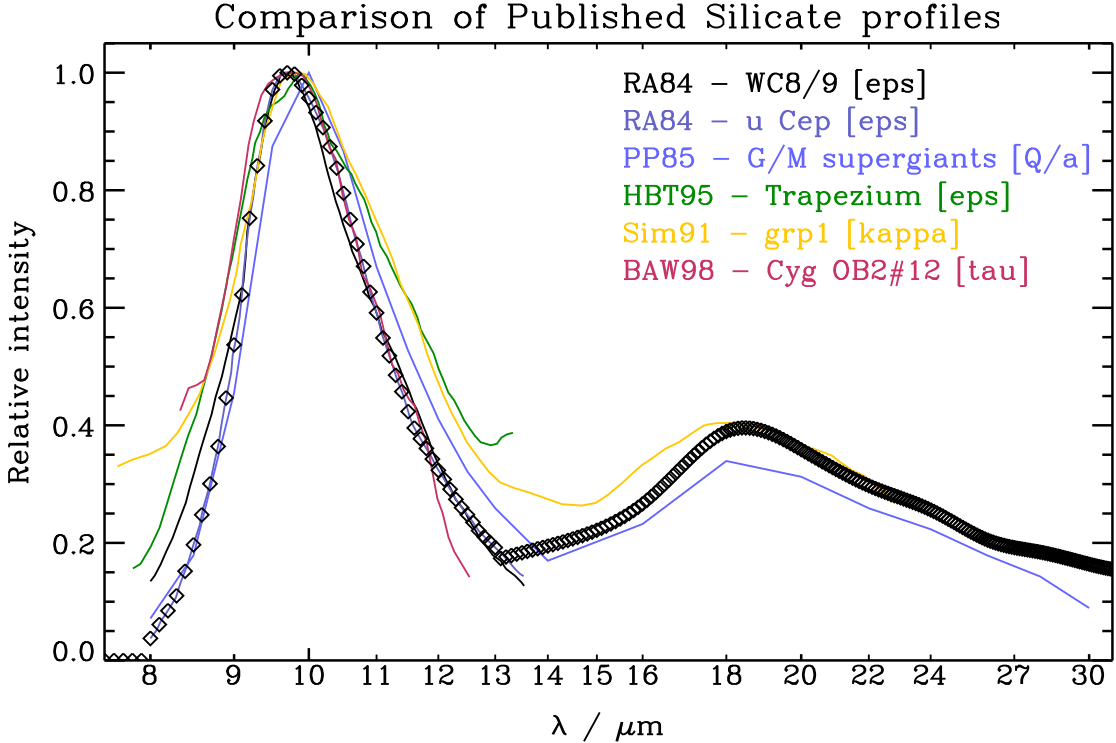}
\caption{Comparison of different published silicate emissivities.  The
adopted emissivity is traced by the black diamonds.}
\end{figure*}

We removed the silicate features from the spectra by using the above
emissivity profiles, $\epsilon(\lambda)$, as follows:
\[ F(\lambda) = f(\lambda) e^{\tausil\epsilon(\lambda)}\]
where $F(\lambda)$ is the corrected spectrum, $f(\lambda)$ is the observed
spectrum, and $\tausil$ is the peak optical depth of the feature.  The best
fitting profile and the peak optical depth were determined subjectively by
requiring that the cleaned spectrum be continuous and free of residual
features.  The alternative of using $\chi^2$ methods was considered, but
the errors at the beginning and the end of the profiles produce large
deviations in the corrected spectra, which hide the deviations caused by
more important errors in the profile where the emissivity is large.  This
is particularly evident over the $9.7\mic$ feature, which is strong and
narrow.

The best emissivity profile for the $9.7\mic$ feature is the $\mu$ Cep
profile from RA84, and proper correction for all the sources is achieved
with $\tausil = 2.9 \pm 0.1$.  The weak and sharp feature remaining at
$9.4\mic$ is not real, but it is due to a slight error in the determination
of the emissivity profile in a region of rapid variation and where the
atmospheric absorption is greatest.  A change of 0.1 in $\tausil$ produces
a clearly noticeable effect in the corrected spectra.

The two $18\mic$ emissivity profiles are similar in shape but differ in
strength relative to the $9.7\mic$ feature: P\'egouri\'e and Papoular
(1985) determine $\tau(18)/\tau(9.7) = 0.35$, while Simpson (1991)
constrained her continuum model to produce $\tau(18)/\tau(9.7) = 0.40$.  We
have used the P\'egouri\'e and Papoular (1985) profile multiplied by 1.15
to produce $\tau(18)/ \tau(9.7) = 0.40$, and joined it to the RA84 $\mu$
Cep profile at $13\mic$.  We note that the P\'egouri\'e and Papoular (1985)
profile of the $9.7\mic$ feature is much narrower than the cocoon star
feature.  This combined profile gives a deeper minimum at $12\mic$ than if
we had used the Simpson (1991) profile alone.  The corrected (and
dereddened, see below) spectra, are shown in Figure 8.

Our value of $\tausil$ is significantly higher than the value deduced by
Schutte et al.~(1998), who measured the depth of the feature relative to
the level of the spectrum adjacent the feature.  While their method is fine
for their purpose, which is to intercompare the variation of this and other
features in several sources, their result does not account for the
absorption at 12--$14\mic$, as seen in Figure 8.

We are confident that our method is reliable for the $9.7\mic$ feature, but
attempts to use to SWS spectra to place independent constraints on the
strength and shape of the $18\mic$ feature using the same criteria adopted
for the $9.7\mic$ feature proved unsuccessful.  The reason is probably that
beyond 18--$20\mic$ these spectra are a blend of various cocoon stars and
of the Pistol \Hii\ region, and are thus intrinsically too complicated for
a subjective determination of a ``smooth'' spectrum.  Furthermore, the task
was intrinsically more difficult due to the weaker and broader nature of
the $18\mic$ feature.

\subsection{The Extinction Law}

\begin{figure*}    
\centering \leavevmode 
\epsfxsize=0.83\textwidth\epsfbox{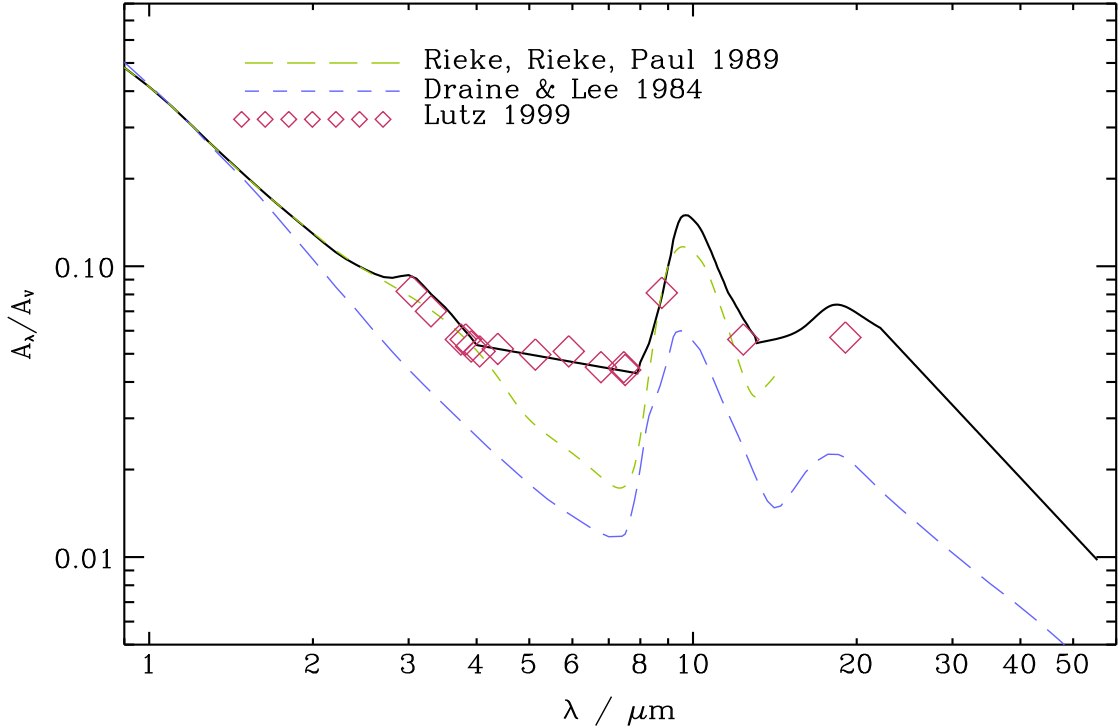}
\caption{Adopted extinction law compared to Draine \& Lee (1984), Rieke,
Rieke, \& Paul (1989), and Lutz (1999).}
\end{figure*}

The extinction law, $A_\lambda/\Av$, toward the GC is known to give rise to
more extinction than the standard law both in the silicate bands and in the
5--$8\mic$ spectral range: Roche \&\ Aitken (1985, RA85) have shown that
$\Av/\tausil\simeq 8.5$, or $\Ak/\tausil\simeq 0.70$ with their extinction
law (van de Hulst no.~15), toward the GC, which is about half the value in
the solar neighborhood; and Rieke, Rieke, \& Paul (1989) have shown on
the basis of the \Brg\ to \Bra\ ratios, that the extinction at $4\mic$
towards the GC is higher than expected with standard laws.  This results
was confirmed and extended by Lutz (1999) using hydrogen emission lines
throughout the 2.5 to $12\mic$ range.

We have combined the above results to construct a reddening law appropriate
for the GC as follows:  \begin{enumerate}
\item shortward of the $2.5\mic$ we follow the Rieke \& Lebofsky (1985) law
which was determined for GC sources.  At $0.9\mic < \lambda < 2.5\mic$ this
law can be approximated by a power law $A_\lambda\sim \lambda^{-\alpha}$,
with $\alpha = 1.63$;
\item between 2.5 and $3.8\mic$ some structure is seen in the Lutz (1999)
results which is due to the ice feature at $2.9\mic$ and to the amorphous
carbon feature at $3.4\mic$.  We modeled these features as Gaussian;
\item at $\lambda \simeq 3.8$--$8\mic$, a power law with $\alpha = 0.35$,
was used to approximate the results of Lutz (1999).  Note that the point at
$6\mic$, where the deviation from the power law is largest, suffers extra
extinction due to the water-ice feature.  At the position of \Bra\ this
law gives the same results as the Rieke, Rieke, \&\ Paul (1989);
\item between 8 and $24\mic$, the silicate absorption features are modeled
using the emissivity adopted in the previous section, with $\tausil = 2.9$;
and
\item beyond $24\mic$ a power law with $\alpha = 2$ is adopted (e.g.~Draine
\&\ Lee 1984).  In practice this last portion of the reddening law does not
affect our results.  
\end{enumerate} 
The adopted extinction law is shown in Figure 9, where other commonly used
reddening laws are shown for comparison, and it is tabulated in Table 3.
With this law, $\Av/\Ak = 8.9$, $\Av/\tausil = 10$, and $\Ak/\tausil =
1.14$.

\begin{table}
\caption{The extinction law}
\leavevmode  \footnotesize
\begin{tabular}{rrrrr}
\hline \\ [-5pt]
$\lambda/\mic$ & $A_\lambda/\Av$ & ~ & $\lambda/\mic$ & $A_\lambda/\Av$ \\
\hline \\ [-5pt]
0.55 & 1.0000 & ~ &  12.00 & 0.0721 \\	
0.65 & 0.7689 & ~ &  12.50 & 0.0634 \\	
0.80 & 0.5584 & ~ &  13.00 & 0.0569 \\	
1.00 & 0.4097 & ~ &  13.50 & 0.0545 \\	
1.25 & 0.2804 & ~ &  14.00 & 0.0553 \\	
1.65 & 0.1736 & ~ &  14.50 & 0.0561 \\	
2.20 & 0.1108 & ~ &  15.00 & 0.0573 \\	
2.50 & 0.0964 & ~ &  15.50 & 0.0589 \\	
3.00 & 0.0923 & ~ &  16.00 & 0.0612 \\	
3.50 & 0.0715 & ~ &  16.50 & 0.0643 \\	
4.00 & 0.0530 & ~ &  17.00 & 0.0677 \\	
4.50 & 0.0510 & ~ &  17.50 & 0.0707 \\	
5.00 & 0.0492 & ~ &  18.00 & 0.0726 \\	
5.50 & 0.0477 & ~ &  18.50 & 0.0730 \\	
6.00 & 0.0464 & ~ &  19.00 & 0.0721 \\	
6.50 & 0.0452 & ~ &  19.50 & 0.0704 \\	
7.00 & 0.0441 & ~ &  20.00 & 0.0683 \\	
7.50 & 0.0431 & ~ &  20.50 & 0.0663 \\	
8.00 & 0.0463 & ~ &  21.00 & 0.0644 \\	
8.50 & 0.0628 & ~ &  21.50 & 0.0626 \\	
9.00 & 0.0989 & ~ &  22.00 & 0.0610 \\	
9.50 & 0.1454 & ~ &  22.50 & 0.0586 \\	
10.00 & 0.1431 & ~ & 23.00 & 0.0561 \\   
10.50 & 0.1250 & ~ & 23.50 & 0.0537 \\   
11.00 & 0.1023 & ~ & 24.00 & 0.0515 \\   
11.50 & 0.0835 & ~ &$>24.0$& $\propto \lambda^{-2}$ \\
\hline \end{tabular} \end{table}

\subsection{Extension of the Cocoon Stars}

\begin{figure*}[bht]       
\plottwo{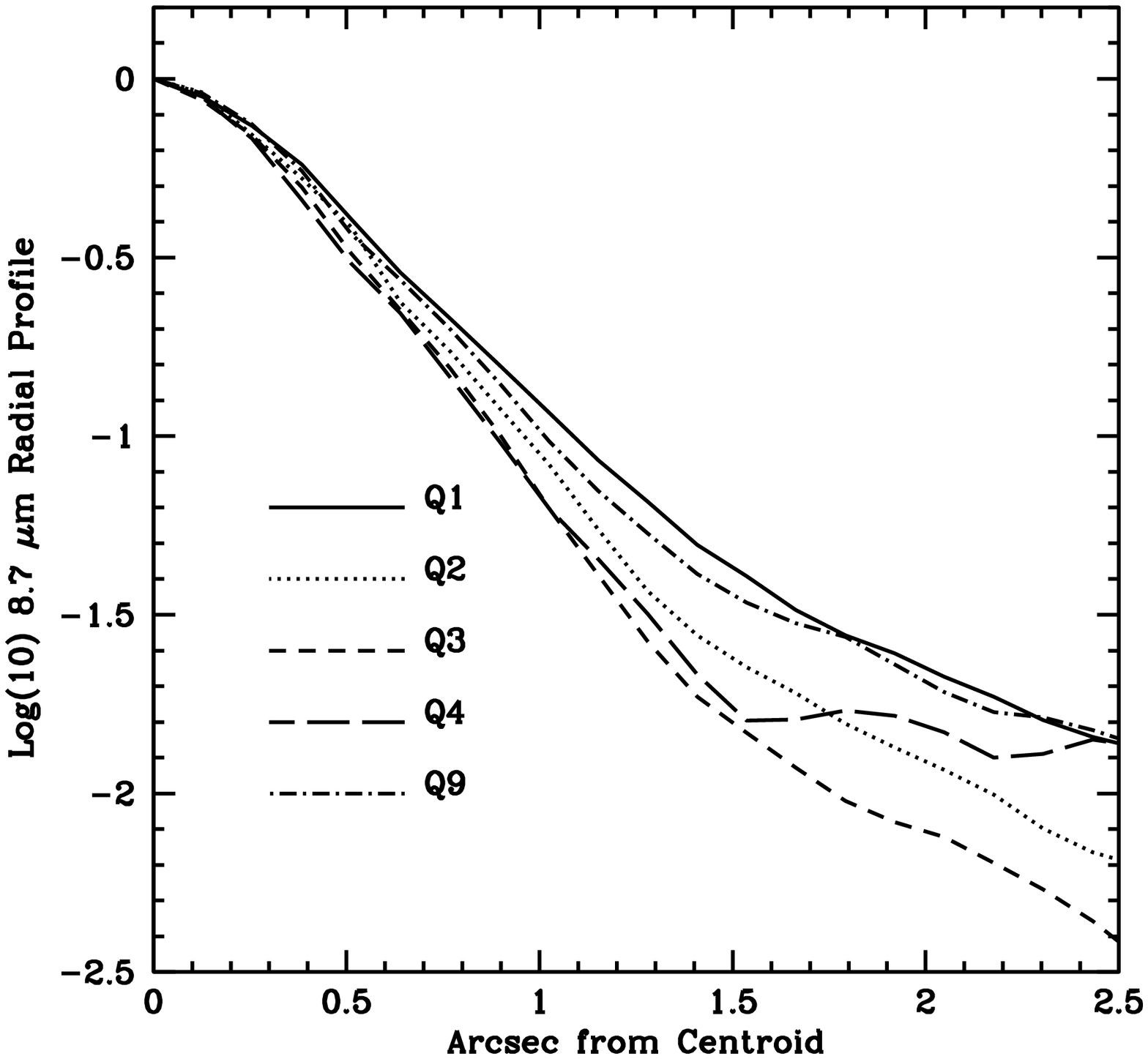}{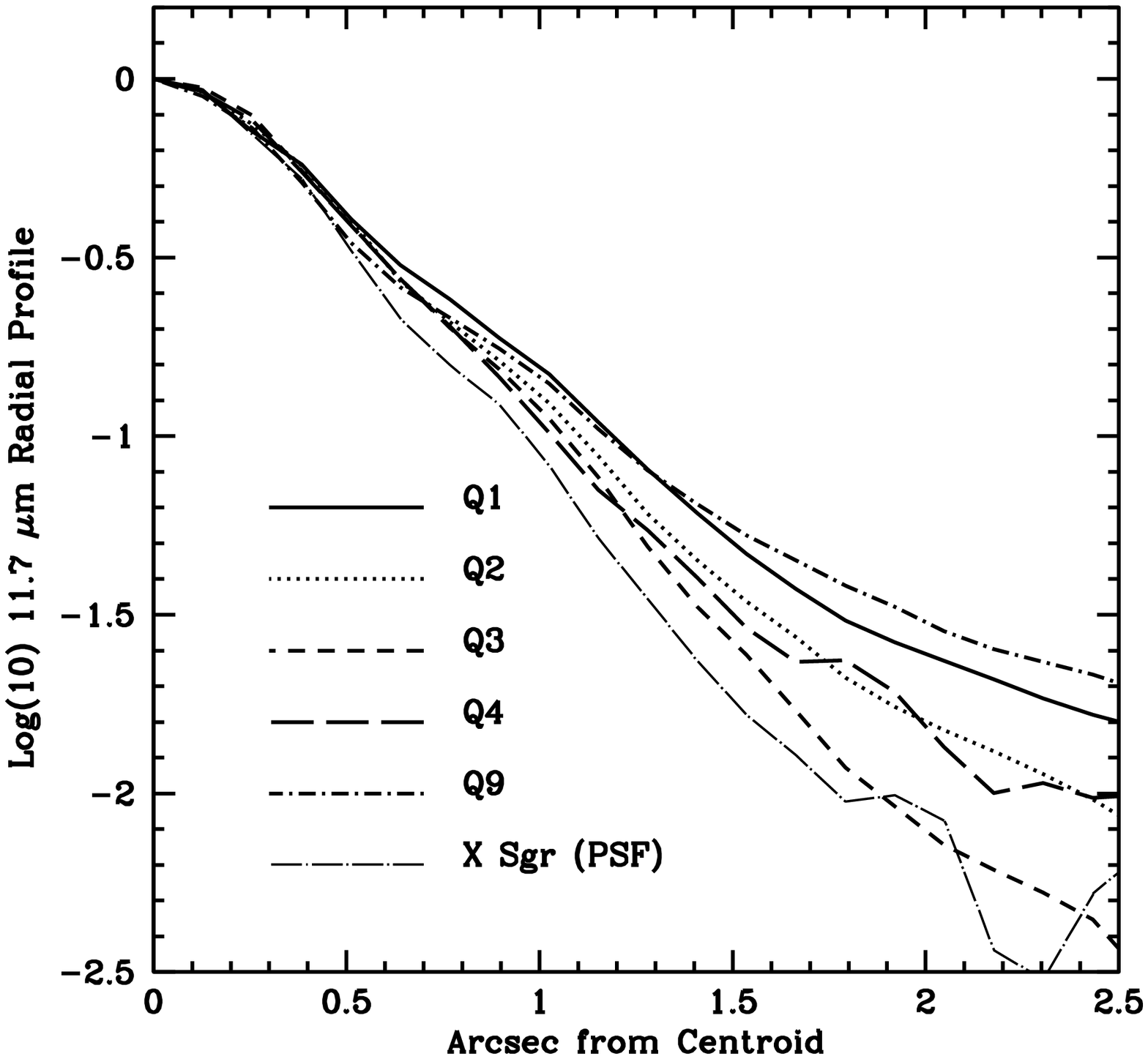}
\caption{Radial profiles of the cocoon stars at 8.7 and $11.7\mic$.  The
profile of the reference star X Sgr at $12\mic$ is also shown for
comparison.  All observations were obtained at airmass $\simeq 2.2$. }
\end{figure*}

Inspection of the Palomar images (Figure 4) shows significant extended
emission around some of the cocoon stars.  Figure 10 shows the radial
profiles which were measured on a special mosaic composed from selected
exposures of superior seeing obtained during a single night at each
wavelength.  The radial profile of the isolated, but rather faint star, X
Sgr, that was observed at nearly the same airmass, is also shown for
comparison (the profile of $\alpha$ Her observed at unit airmass, not
shown, is even narrower, with a ${\rm FWHM} =0.69''$ vs.~$0.59''$ for
the diffraction size of $1.22 \lambda/D$).

The profiles of Q3 and of Q4 (out to $1.5''$, beyond which Q4 is likely
to be contaminated by the bright sources nearby) are similar to that of
reference star.  Q1 and Q9 show wings that are almost $10\times$ higher,
while Q2 shows a weaker, but clear excess.  The profiles shown are
azimuthal averages, but the extensions are clearly asymmetric and do not
show a common preferred direction, as seen in Figure 4.

Could the observed profile wings be due to other nearby red stars?  The
high spatial resolution $2.05\mic$ (F205W) HST/NICMOS images of Figer et
al.~(1999) show three reddish stars close to Q1, a large blob of diffuse
emission NW of Q2, weak but not red stars near Q3 and Q4, and a fairly
isolated Q9.  The blob near Q2, the brightest of the group, is likely to be
an artifact.  We have performed photometry of all the ``companions'', and
find that the red ones are far less red than the cocoon stars, and simple
extrapolations of their energy distributions to 8 and $12\mic$ would
produce negligible flux at those wavelengths.  Also, we note that Q9, which
is the most extended cocoon star, is also the most isolated in the near IR.
We therefore conclude that Q1, Q9, and to a lesser extend Q2 are
intrinsically extended.  At the distance of the Galactic Centre (8 kpc), an
angular radius of $2''$ corresponds to a physical radius of 0.08 pc or
16,000 AUs (radius to zero power).  The true radius should be somewhat
smaller, given the seeing of $\simeq 1''$ for the observations, but will
remain nevertheless very important.

\subsection{Energy Distributions and Luminosities}

Using the reddening law constructed above we dereddened our spectra
assuming $\Ak = 3.3\,$mag ($\Av = 29\,$mag), as determined by Figer et
al. (1998) from the average color excesses of the early-type stars in the
cluster.  The uncertainty in $\Av$ is 5 mag.  The results are shown in
Figure 11 together with the dereddened ground-based data of FMM99 and of
Okuda et al.~(1990)

\begin{figure*}    
\centering \leavevmode 
\epsfxsize=0.95\textwidth\epsfbox{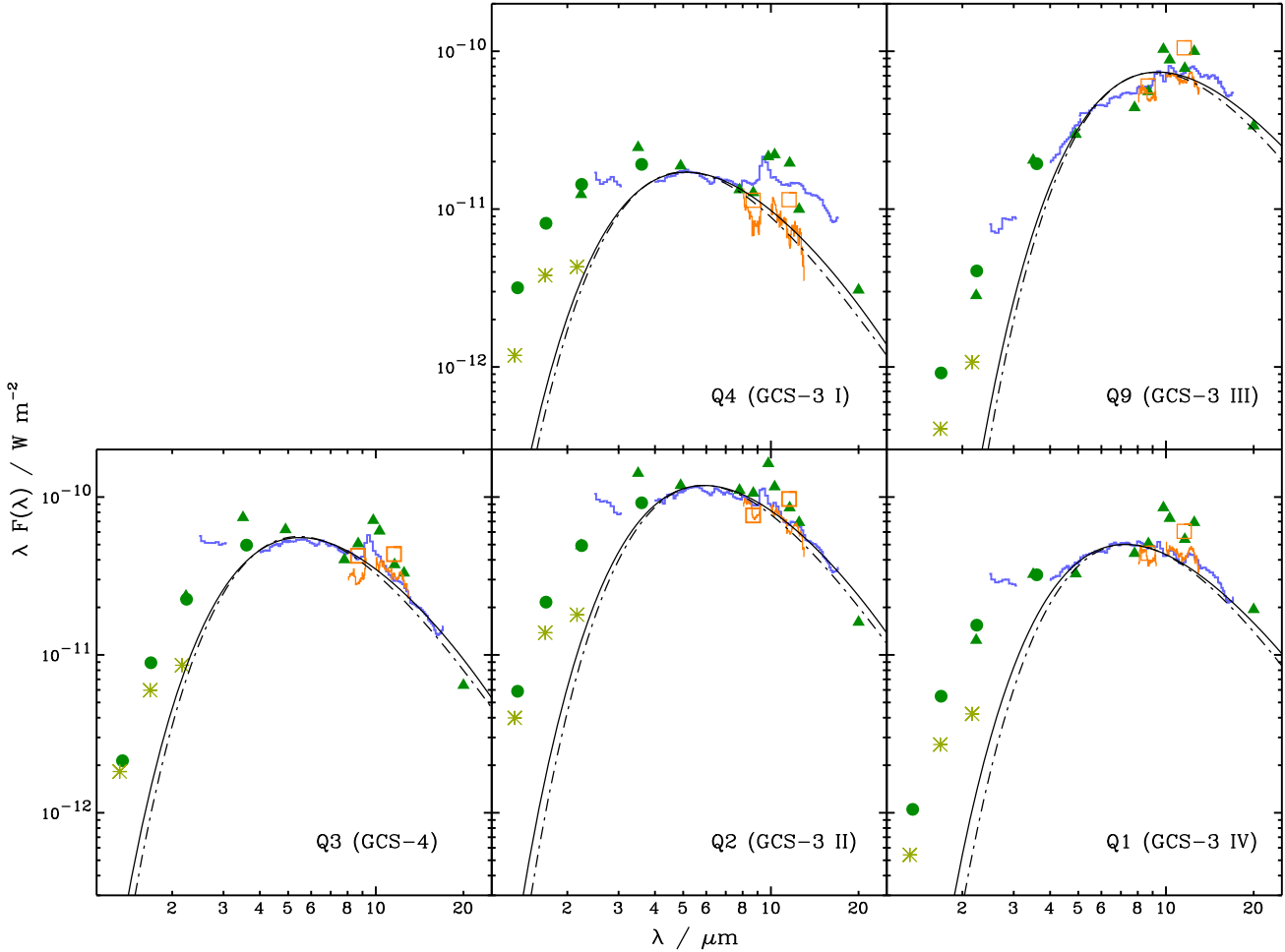}
\caption{Dereddened spectral energy distributions of cocoon stars; symbols
are as in Figure 5.  The models shown are simple blackbodies (dot-dash
lines) and DWCL-like shells (solid lines).  }
\end{figure*}

The resulting SEDs were modeled in three ways (see also Figure 11):
\begin{enumerate}
\item a constant temperature, optically thick shell, modeled by a Planck
function (dot-dash lines in Figure 11);
\item a constant temperature, optically thin shell, modeled by a Planck
function multiplied by a $\lambda ^{-1}$ emissivity law: this produces an
energy distribution narrower than the optically thick models (not shown);
\item an optically thin but geometrically thick dust shell with $r^{-2}$
density law (constant velocity wind) as used by WHT to model known DWCL
stars.  The shape of the model SED is determined by the dust temperature,
$T_0$, at the inner radius of the shell, $R_0$, and the shell thickness,
$H$, which is measured in units of $R_0$.  A dust emissivity proportional
to $\lambda ^{-1}$ was used.  This model produces an SED that is wider than
a Planck function modified with the same emissivity law, and whose width
increases as the physical thickness increases (solid line in Figure 11).
\end{enumerate}
More detailed models of the DWCL dust shells have been computed by Zubko et
al.~(1998).  While those models provide useful new results on the
properties of the dust, the SEDs they produce are not significantly
different from those of WHT, so we will stay with the WHT models for the
remainder of this work.  

The models described above were normalized to the data at $\lambda$ between
6.5 and $7\mic$, where the spectrum is reasonably free of dust or ice
absorption features.  The model parameters are listed in Table 4.  In all
cases the shell thickness is very large, and increasing $H$ further does
not have a significant effect on the SED, thus these models only place a
lower limit to the shell thickness.  A change of 20 K in $T_0$ produces a
significant change in the SED, and the value of 20 K can be taken as the
uncertainty in $T_0$.

\begin{table}
\caption{Model parameters and results} 
\leavevmode \footnotesize
\begin{tabular}{lccrc}  
\hline \\ [-5pt]
Source  & $T_{\rm BB}$ & $T_0/K$  &  $H$~ &  $\log(L/\Lsun)$ \\ [2pt]
\hline \\ [-5pt]
Q1  & 500 & 550 & 100~& 4.5 \\
Q2  & 625 & 675 &  50~& 4.9 \\
Q3  & 650 & 700 &  50~& 4.6 \\
Q4  & 725 & 775 &  50~& 4.1 \\
Q9  & 400 & 425 & 100~& 4.7 \\[2pt]
\hline 
\end{tabular} 
\end{table}

As can be seen in the Figure 11, the DWCL-like models can reproduce quite
well the SEDs over the 4--$17\mic$ range.  The models reproduce best the
SEDs for Q2, which is the brightest in the mid-IR, and Q3, which is the
most isolated source.  Q1 is not very bright at long $\lambda$, and its
spectrophotometry is likely to suffer some contamination from the nearby
Q2, especially from the first Airy ring of the PSF of the latter.  Q4 also
suffers heavy contamination by the nearby sources, as described in \S
2.5.  Q9 is the the reddest source, its SED is somewhat peculiar, possibly
due to some contamination at short $\lambda$; for this source the model is
mediocre at best.  

Below $\simeq 4\mic$ the model SEDs fall far below the observations,
suggesting that the actual hot dust contribution is larger than in the
model.  Luminosities were computed integrating underneath the model SEDs;
they are $10^{4.1}$ to $10^{4.9}\Lsun$ for the DWCL-like models, while the
blackbody models give luminosities about 0.1 dex lower (see Table 4).
These values do not differ significantly from previous determinations.


\section{Discussion}
\subsection{The Origin of the Absorption Features}

Combining the adopted value of $\Av$ with the value of $\tausil = 2.9\pm
0.1$ determined in \S 3.2, gives $\Av/\tausil \simeq 10\pm 2$.  This,
within the uncertainties, is equal to the nominal value of 8.5 determined
by RA85 for the sources near Sgr A$^*$.  This implies that all the silicate
absorption, and by extension all the extinction, takes place along the line
of sight.  This is not surprising: the infrared polarisation toward the
cocoon stars and toward Sgr A$^*$ is about the same both in amount and in
position angle (Kobayashi et al.~1983), and the depths of the other ice and
gas features are all similar to, though usually slightly lower than, their
corresponding values toward Sgr A$^*$ (Gerakines et al.~1999, Chiar et
al.~2000, Moneti \& Cernicharo 2000).  The general conclusion is that there
is slightly less absorbing material toward the Quintuplet Cluster than
toward Sgr A$^*$, and also that the fraction of cold molecular cloud
material is lower toward the cluster.  All this speaks against any
absorption being produced by the cocoon stars themselves, and supports the
conclusion that all the absorption features observed are of interstellar
origin.

The lack of intrinsic absorption speaks against the interpretation of the
cocoon stars as deeply embedded YSOs, where the deep silicate absorption is
ascribed to the molecular cloud in which the YSO is deeply embedded.
Furthermore, the $9.7\mic$ feature toward deeply embedded YSOs is usually
broader than that of the ISM, and is better reproduced by the Trapezium
profile than by the $\mu$ Cep profile.  Also, the spectra of embedded YSOs
usually show warm CO and H$_2$O gas in absorption (see, e.g.~van Dishoeck
\& Helmich 1996), presumably arising in a region where this gas is warmed
either by the young stars itself or by shocks produced when the wind and/or
the outflow from a young stars strikes the surrounding molecular material.
No such warm gas is found toward the cocoon stars.  Indeed, molecular line
studies have shown the Quintuplet Cluster is located in a region free of
molecular gas (Serabyn and G\"usten 1991).

\subsection{On the DWCL Hypothesis}

The luminosities derived in \S 3.5 are those of the shells.  The total
stellar luminosity, assuming that the shell absorbs and emits
isotropically, will be the sum of the observed (and dereddened) shell
luminosity and the observed (and dereddened) stellar luminosity, the latter
being that fraction of the stellar luminosity not absorbed by the shell,
and transmitted toward the observer.  In the following we will alway refer
to these observational quantities.   

For most of the DWCL stars studied by WHT, the ratio of shell to stellar
luminosity is $\lesssim 0.1$, while only for the most extreme 20\% it is
between 0.5 and 0.7.  Assuming the lower value, i.e. $L_{\rm shell}/L_{\rm
star} = 0.1$, the total luminosities of the cocoon stars would be
comparable to those of typical Galactic and Magellanic Clouds WCL stars,
$\sim 10^{5.5}\Lsun$.  But the lack of near IR spectral lines (especially
in the $J$-band, see FMM99) suggests that $L_{\rm shell}/L_{\rm star}$ must
be large.  If half the total stellar luminosity were re-emitted by the
shell, and, for the sake of argument, if they were both Planckian with
effective temperatures of 25,000 and 800 K respectively, then the two
energy distributions would intersect at $\simeq 1.37\mic$, and the
$1.25\mic$ continuum of the star would be $\sim 3\times$ that of the shell.
In this situation, the ratio $L_{\rm shell}/L_{\rm star} = 1.0$, already
larger than for the most extreme DWCL stars in WHT.  Increasing the stellar
temperature to 35,000 K will move the intersection point to $1.27\mic$, and
the stellar continuum will be only 20\% higher than the shell continuum.
Under such circumstances the stellar emission line spectrum should still be
easily visible in the $J$-band, unlike the results of FMM99.  Increasing
the shell luminosity to $10\times$ the (transmitted) stellar luminosity
would be necessary for the shell continuum to dominate the stellar
continuum at $1.25\mic$.  Thus, for the $J$-band spectrum to be dominated
by the shell, the shell luminosity would have to be $\moresim 90\%$ of the
total (shell + star), and the star has to be very hot.  In this case, the
UV optical thickness of the shells would have to be much greater than
unity, which would probably be inconsistent with the optically thin shell
in the near and mid IR, but it would bring the total luminosities roughly
equal to the observed luminosities, and these values are rather low for
typical DWCL stars.

In summary, if the shells reprocess only a small part of the true stellar
luminosity, then the true stellar luminosity is comparable to that of known
DWCL stars, but the stellar spectrum should be observable in the $J$-band.
If, on the other hand, the shells reprocess most of the luminosity of the
central star, in which case our models are likely to fail on optical
thickness grounds, our models predict a total luminosity much lower than
known DWCL stars.

The physical size of the shells can be estimated from their luminosity and
$\Tbb$.  Taking typical values of $L = 50,000\Lsun$ and $\Tbb = 500K$, we
derive a physical diameter of 250 AU, which corresponds to and angular
diameter of 35 milliarcsec.  This value should correspond roughly to the
inner diameter of the shell.  The outer diameter will be 50--$100\times$
larger, or $\simeq 2$--$3''$.  These values are very rough, but they are
overall consistent with the observed extensions observed in the Palomar
images, though the coincidence could be fortuitous.

\subsection{Other Interpretations}

Glass et al.~(1999) suggested that the cocoon stars could be
self-enshrouded massive stars of the type suggested by Bernasconi \& Maeder
(1996).  Without discussing those models, we note that the measured shell
luminosities in Table 4, if they represent the total stellar luminosity,
correspond to stars of initial mass between 15 and $25\Msun$ in the
theoretical HR diagrams of Bernasconi \& Maeder (1996).  This would place
them below the $\simeq 40\Msun$ limit of the stars that would remain
obscured on the Main Sequence if those models were applicable.  Secondly,
if the cocoon stars are coeval with the rest of the cluster, their age of
$\simeq 4\,$Myrs would make them too old for this proposed evolutionary
stage.

Okuda et al.~(1990) already discussed why other evolved stars are unlikely
explanations of the nature of the cocoon stars.  We add here that the
luminosities we derived are somewhat too high for AGB stars, and also that
the cluster is too young for AGB stars to have formed: stars much more
massive than $\sim 8\Msun$ are still on the main sequence (FMM99).  


\section{Conclusions}

New mid-infrared spectroscopic observations of the enigmatic cocoon stars
in the Quintuplet Cluster were obtained in order to put further constraints
on their nature.  The high spectral/low spatial resolution spectroscopy
did not reveal any feature that could be assigned to the cocoon stars
themselves.  The low spectral/high spatial resolution spectrophotometry was
used to study the form of the silicate absorption in these stars, and we
conclude that (i) the depth of the absorption is the same for all the
stars, and (ii) there is no evidence for any silicate emission or
absorption intrinsic to the cocoon stars.  In fact, all of the absorption
features are consistent with an  origin in the molecular clouds and in the
more tenuous ISM along the line of sight.  

We have examined carefully the hypothesis that the cocoon stars are extreme
DWCL stars.  The SEDs of the cocoon stars can be reproduced by the models
of DWCL shells presented by WHT, but the high UV optical thickness needed
to reprocess most of the luminosity of the central star suggests that these
models might not be applicable to such extreme stars.  Furthermore, the
luminosities derived are significantly lower than those of known DWCL
stars, and also these stars are extremely rare in the Galaxy, though they
are concentrated in the GC.  Yet, stars of the carbon sequence are
necessary to produce featureless spectra.

The new observations do not provide any new constraints in favour of any of
the proposed identifications of the cocoon stars but they do provide new
arguments against all three of the main hypotheses that have generally been
considered, namely dust-enshrouded youg stars, late-type (OH/IR or AGB)
stars, or late-type WC stars.  The nature of these stars remains an enigma.


\section*{Acknowledgments}

We would like to acknowledge T.~Hayward and T.~Herter for assistance in
obtaining and reducing the SpectroCam-10 data and Cornell University for
providing access to the Hale telescope.  S.R.S. acknowledges support by
NASA grant NAG2-207 and NSF grant AST-9218038.  NASA grants NAGW-2551 and
NAGW-2870 supported the development of SpectroCam-10 and the detector.


\end{document}